\documentclass{llncs}

\usepackage{amsfonts,amsmath,amssymb}
\usepackage[noend]{algorithmic}
\usepackage{algorithm}
\usepackage{graphicx}

\usepackage{listings}

\usepackage{multirow}

\lstdefinelanguage{pseudocode}{
morekeywords={if, then, else, for, return, in, while, case, do, forever, foreach, False, True},
morecomment=[l]\#,%
morecomment=[l]//,%
moredelim=[il][\it]{|},
moredelim=[il][]{[},escapeinside={(*@}{@*)}
}

\begin{document}

\title{Efficient chaining of seeds in ordered trees}

\author{Julien Allali\inst{1,2,3}, C\'edric Chauve\inst{3},
   Pascal Ferraro\inst{1,2,4} and Anne-Laure Gaillard\inst{1}}

\institute{LaBRI, Universit\'e Bordeaux 1, IPB, CNRS
   \email{[julien.allali@labri.fr, anne-laure.gaillard@labri.fr, pascal.ferraro@labri.fr]}
\and
Pacific Institute for Mathematical Sciences and CNRS UMI3069
\and 
Department of Mathematics, Simon Fraser University
\email{cedric.chauve@sfu.ca}
\and
Department of Computer Science, University of Calgary}

\maketitle{}

\begin{abstract} 
  We consider here the problem of chaining seeds in ordered
  trees. Seeds are mappings between two trees $Q$ and $T$ and a chain
  is a subset of non overlapping seeds that is consistent with respect
  to postfix order and ancestrality. This problem is a natural
  extension of a similar problem for sequences, and has applications
  in computational biology, such as mining a database of RNA secondary
  structures.  For the chaining problem with a set of $m$ constant
  size seeds, we describe an algorithm with complexity $O(m^2\log(m))$
  in time and $O(m^2)$ in space.
\end{abstract}

\section{Introduction}
\label{sec:introduction}

Comparing sequences is a basic task in computational biology, either
for mining genomics database, or for filtering large sequence
datasets. The exponential increase of available sequence data
motivates the need for very efficient sequence comparison algorithms.
A fundamental application of sequence comparison is to search
efficiently in a database a set of sequences close to a query
sequence. Indeed, the pairwise comparison between the query and every
sequence of the database cannot practically be applied due to the quadratic time
complexity of edit distance computation. A typical approach to tackle
this issue is to rely on short sequences, called {\em seeds}, present
in the query. These seeds can be detected very quickly in the
database using indexing techniques, then a maximal set of seeds, called a {\em chain},
that tiles both the query and a sequence of the database, 
must be identified while conserving the same order in both sequences.
Widely used programs such as
BLAST~\cite{Altschul-blast} and
FASTA~\cite{Lipman-fasta,Pearson-fasta} rely on such an
approach. We refer the reader
to~\cite{Aluru-handbook,Gusfield-algorithms} for surveys of sequence
comparison in computational biology. From an algorithmic point of view, an optimal chain between
two sequences, given $m$ seeds, can be computed in $O(m \log(m))$ time
and $O(m)$ space~\cite{Joseph-determining}
(see~\cite{Ohlebusch-chaining} for a recent survey).

With the recent development of high-throughput genome annotation
methods, similar problems appear to be relevant for the analysis of
more complex biological structures
. For instance, RNA
secondary structures can be modeled by a tree or a graph whose nodes
are the nucleotides and whose edges are the chemical bonds between
them \cite{sha90}.  Large databases have been constituted for this
kind of biological data, such as Rfam~\cite{RFAM}. Comparing and
mining large RNA secondary structure databases is now an important
computational biology problem. The initial approach to these problems
relied on extensions of the notion of edit distance to
ordered trees, pioneered by Zhang and Shasha~\cite{Zhang-simple}. The
tree edit approach has been extended in several ways since then,
leading either to hard problems, when a comprehensive set of edit
operations is considered~\cite{Jiang-general}, or to algorithms with a
time complexity, at best cubic, even with a minimal set of edit
operations~\cite{Weimann-optimal,Zhang-simple}.

Recently, 
Heyne {\it et al.}  \cite{Heyne-lightweight} introduced a chaining
problem on another representation of ordered trees called arc
annotated sequences, that they solved using dynamic programming. Their
seeds are {\em exact common patterns} and they applied their algorithm
for RNA secondary structure comparison: once a maximal chain of seeds
between two given RNA secondary structures is detected, the regions
between successive seeds are processed independently using an edit
distance algorithm, which speeds up significantly the comparison
process. From what we know so far, \cite{Heyne-lightweight} is the
first paper addressing a chaining problem in trees. However, when
applied for chaining seeds in sequences, their algorithm complexity is
higher than the best-known chaining algorithms for sequences, which
raises the question of a more efficient algorithm, of both
theoretical and practical interest.

After some preliminaries (Section~\ref{sec:preliminaries} and
\ref{sec:2D-preliminaires}), we describe in Section~\ref{sec:2D} an
algorithm for finding the score of a maximum-score chain between two
ordered trees (Maximal Chaining Problem) in $O(m^2 \log(m))$ time and
$O(m^2)$ space 
when there are $m$ seeds of constant size,
thus improving on the result of Heyne {\it et
  al.}~\cite{Heyne-lightweight}. We conclude with further research
avenues.

\section{Background and problem statement}
\label{sec:preliminaries}

Let $T$ be an ordered rooted tree of size $n$. Nodes of $T$ are
identified with their postfix-order index from 0 to $n-1$.  Thus,
$n-1$ represents the root of $T$.  $T_{i}$ is the subtree of $T$
rooted at $i$. We denote by $T[i,j]$ the forest induced by the nodes
that belong to the interval $[i,j]$; if $i>j$, then $T[i,j]$ is
empty. The partial relationship ``$i$ is an ancestor of $j$'' is
denoted by $i\prec j$.  For a tree $T$ and a node $i$ of $T$, the
first leaf visited during a postfix traversal of $T_i$ is denoted by
$l(i)$ and called the {\em leftmost leaf} of the node $i$.  The
ordered forest induced by the proper descendants of $i$ is denoted by
$\widehat{T_i} = T[l(i),i-1]$.

\begin{definition}\em
  \label{def:internal_tree} Let $T$ be an ordered rooted tree:
  \begin{enumerate}
  \item
    Let $G=\{g_{0},\dots,g_{k-1}\}$ be an ordered set of $k$ nodes of $T$, with
    $0\le g_j <n$.  If the subgraph of $T$
    induced by $G$ is connected, then $G$ is called an {\em internal tree}
    rooted at $g_{k-1}$ also referred as $r_G$.
  \item The set of leaves of the internal tree $G$ is denoted by $L(G)$.
  \item A node $g_j$ of $G$ is said to be {\em completely inside $G$}
    if $g_j$ is not a leaf of $T$ and all its children belong to $G$. The
    set of nodes of $G$ that are not completely inside $G$ is called
    the {\em border of $G$} and is denoted by $B(G)$.
  \item Two internal trees $G^1$ and $G^2$ {\em overlap} if they share
    at least one node, {\em i.e.} $G^1 \cap G^2 \not= \emptyset$.
  \end{enumerate}
\end{definition}

We now recall the central notion of {\em valid mapping} between two trees
introduced in~\cite{sha90}
for the tree edit distance.  Given two trees $Q$ and $T$, a valid
mapping $P$ between $Q$ and $T$ is a set of pairs of
$Q\times T$ such that, if $(q_i,t_i)$ and $(q_j,t_j)$ belong to
$P$, then
    \begin{enumerate}
    \item  $q_i=q_j$ if and only if $t_i=t_j$,
    \item  $q_i<q_j$ if and only if $t_i<t_j$,
    \item  $q_i\prec q_j$ if and only if $t_i\prec t_j$.
    \end{enumerate}
    In the following we use the term \emph{mapping} to refer a
    valid mapping.
    Given a mapping $P$ between $Q$ and $T$, the smallest
    internal tree of $Q$ (resp. $T$) that contains all nodes of $Q$
    (resp. $T$) belonging to a pair of $P$ is denoted by
    $Q_P$ (resp.  $T_P$). $Q_P$ and
    $T_P$ are respectively called the internal trees of $Q$
    and $T$ induced by $P$. 

\begin{definition}
  \label{def:seed}\em
  Let $Q$ and $T$ be two ordered trees. 
  \begin{enumerate}
  \item A \emph{seed} $\mathcal{M}$ between $Q$ and $T$ is a mapping
    between $Q$ and $T$ such that
    $(r_{Q_\mathcal{M}},r_{T_\mathcal{M}}) \in \mathcal{M}$ and all
    the nodes of the border of $Q_\mathcal{M}$ (resp. $T_\mathcal{M}$)
    belong to a pair of $\mathcal{M}$.
  \item The border (resp. leaves) $B(\mathcal{M})$
    (resp. $L(\mathcal{M})$) of the seed $\mathcal{M}$ is the set of
    pairs $(x,y)\in \mathcal{M}$ such that $x\in B(Q_\mathcal{M})$ and
    $y \in B(T_\mathcal{M})$ (resp. $x\in L(Q_\mathcal{M})$ and $y \in
    L(T_\mathcal{M})$).
  \item The {\em size} $|\mathcal{M}|$ of the seed $\mathcal{M}$ is
    the number of pairs its mapping contains.
  \item For a set $S$ of seeds, $\|S\|$ is the sum of the sizes of the
    $|S|$ seeds in $S$.
\end{enumerate}
\end{definition}

\begin{definition} 
  \label{def:chainable}
  \label{def:2D-chain}\em
  Let $Q$ and $T$ be two ordered trees.
  \begin{enumerate}
  \item
    A pair $(P^1,P^2)$ of seeds between
    $Q$ and $T$ is {\em
      chainable} if $Q_{P^1}$ does not overlap $Q_{P^2}$, $T_{P^1}$
    does not overlap $T_{P^2}$, and $P^1 \cup P^2$ is a mapping.
  \item
    A \emph{chain} is a set $C=\{P^0,P^1,\ldots,P^{\ell-1}\}$ of seeds
    between $Q$ and $T$ such that any pair $(P^i,P^j)$ of distinct
    seeds in $C$ is chainable.  
  \item Given a scoring function $v$ for the seeds $P^i$, the score of
    a chain $C$ is the sum of the scores of its seeds: $v(C)=\sum_i
    v(P^i)$.
  \item Given a set $S$ of possibly overlapping seeds between $Q$ and
    $T$, $\mathcal{C}_S(Q,T)$ denotes the set of all possible chains
    between $Q$ and $T$ included in $S$.
  \end{enumerate}
\end{definition}



\begin{figure}[!htb]
\begin{center}\includegraphics[width=0.6\linewidth]{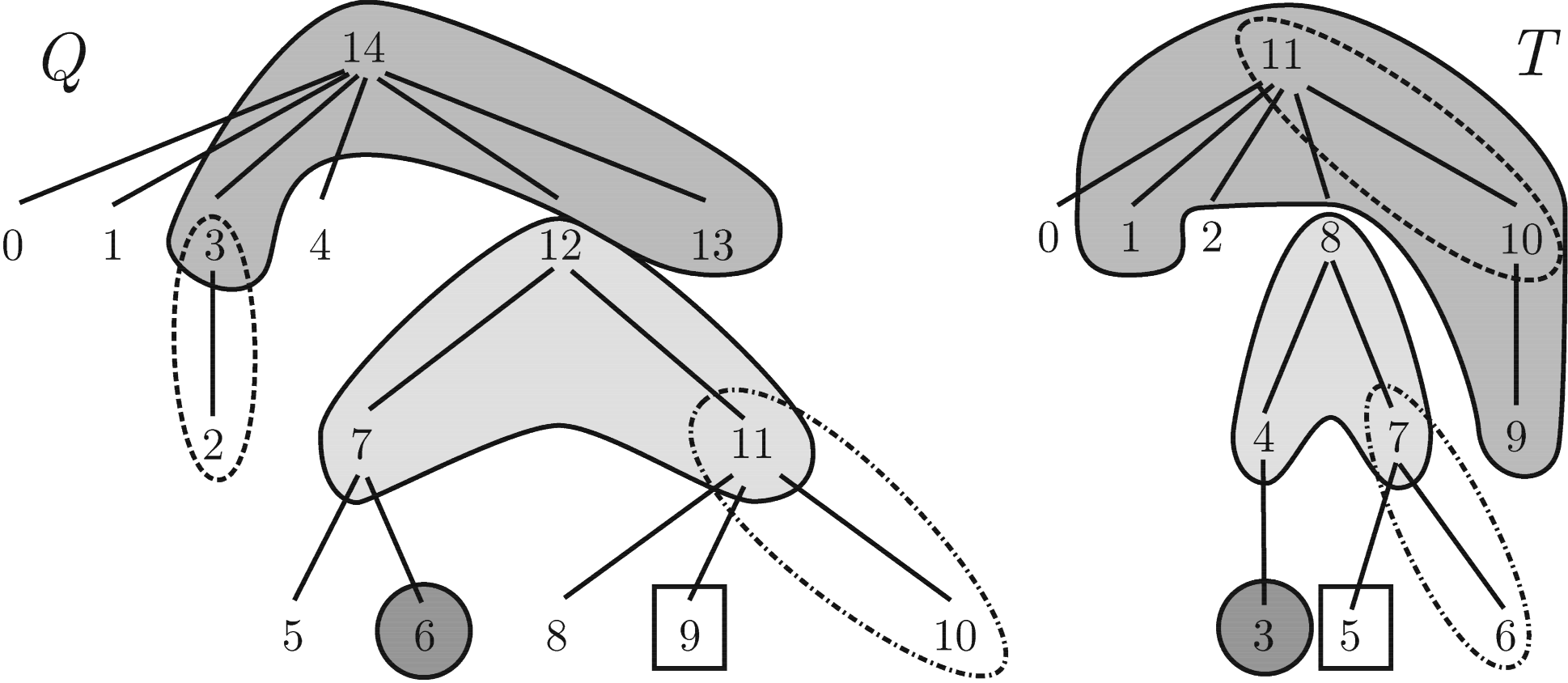}
 \vspace*{-3mm}
\caption{An instance of the MCP problem with 6 seeds:
  $P^0=\{(2,10),(3,11)\}$, $P^1=\{(6,3)\}$, $P^2=\{(9,5)\}$,
  $P^3=\{(10,6),(11,7)\}$, $P^4=\{(7,4),(11,7),(12,8)\}$,
  $P^5=\{(3,1),(13,9),(14,11)\}$. If $v(P^i)=|P^i|$ for every
  seed, an optimal chain is composed of $\{P^1,P^2,P^4,P^5\}$ and has
  score $8$.}
\label{fig:example}
\end{center}
\vspace*{-5mm}
\end{figure}

\noindent{\bf Problem.}
  Maximum Chaining Problem (MCP): 
  \\ 
  Input: A pair $(Q,T)$ of
  ordered rooted trees, a set $S=\{P^0, \dots, P^{m-1}\}$ of $m$ possibly
  overlapping seeds between $Q$ and $T$, and a scoring function $v$ on
  the seeds $P^i$.  
  \\ Output: The maximum score chain $C$ included in $S$.
  $$
  MCP(Q,T,S) = \max\{v(C) ; C\in \mathcal{C}_S(Q,T) \}
  $$

Fig.~\ref{fig:example} shows an instance of the MCP problem with 6 seeds.

\begin{remark}
  \label{rem:forests}
  The notion of mapping extends naturally to ordered
  forests. Hence, if $S$ is a set of seeds between two forests $F_1$
  and $F_2$ such that each seed is a seed between a tree of $F_1$ and
  a tree of $F_2$, then the MCP can naturally be extended to  ordered forests.
\end{remark}


 

\begin{remark}
  \label{rem:trees}
  To compare with chaining algorithms for sequences, we represent a
  sequence $u=(u_0,\ldots,u_{n-1})$ by a unary tree, rooted at a node
  labeled by $u_{n-1}$, where every internal node has a single child
  and $u_0$ is the unique leaf: the sequence of nodes visited by the
  postfix-order traversal of this tree is exactly $u$.
\end{remark}

\paragraph{Motivation and background.}
As far as we know, \cite{Heyne-lightweight} is the only work that
attacks the MCP in tree structures, although the authors describe the
problem in terms of arc-annotated sequences. They proposed a dynamic
programming algorithm to solve the maximum chaining problem with some
restrictions on the seeds (precisely, seeds are maximal exact pattern
common to the considered sequences). This dynamic programming
technique is different from, and in fact simpler than, the approaches
used for the Maximum Chaining Problem in sequences, and, when applied
to arc-annotated sequences with no arc (\emph{i.e.} sequences) it can
be shown this algorithm has a worst-case time complexity in $O(m^2)$,
where $m$ is the number of seeds (see Appendix).

Our main result is an algorithm that solves the Maximum Chaining
Problem with a better complexity than the algorithm
of~\cite{Heyne-lightweight}. After a preprocessing of the $m$ input
seeds of $S$, that can be done in time $O(\|S\|)$ (we discuss in
appendix the complexity of this preprocessing), our algorithm solves
the Maximum Chaining Problem in $O(\|S\|\log(\|S\|) + m\|S\|\log(m))$
time and $O(m\|S\|)$ space. In particular, when applied on sequences
(Remark~\ref{rem:trees}), our algorithm has the same complexity than
the best known algorithms for chaining seeds in sequences
\cite{Ohlebusch-chaining}, that is $O(m\log(m))$ in time and $O(m)$ in
space.

\begin{remark} 
  \label{rem:seeds_order}
  Without lost of generality, from now we assume that the seeds $P^i$
  are sorted increasingly according the postfix number of their roots
  in $Q$, that is: $r_{Q_{P^0}}\le \dots \le r_{Q_{P^i}} \le \dots \le
  r_{Q_{P^{m-1}}}$. For a given chain $C$, the {\em last} seed of $C$
  is then the seed with the highest postfix index in $Q$.
\end{remark}

\section{Combinatorial properties of seeds and chains}
\label{sec:2D-preliminaires}

We first describe combinatorial properties of seeds and chains, that
naturally lead to a recursive scheme to compute a maximum
chain. Indeed, we show that given a chain $C$ and its last seed $P$,
the root and border of $P$ define a partition of both $Q-Q_P$ and
$T-T_P$ into pairs of forests that contain the seeds $C-\{P\}$ and
form sub-chains of $C$. 

\begin{definition}\em\label{def:max}
  Let $P$ be a seed on two trees $Q$ and $T$ and $(a,b;c,d)$ be a
  quadruple such that $l(r_{Q_P})\le a<b<r_{Q_P}$, $l(r_{T_P})\le
  c<d<r_{T_P}$ and the pair of forests $(Q[a,b],T[c,d])$ is free of
  $P$'s nodes ($Q_P\cap Q[a,b]=\emptyset$ and $T_P\cap
  T[c,d]=\emptyset$).
  \begin{enumerate}  
  \item $(a,b;c,d)$ is called {\em $P$-left-maximal} if one of the
    following assertion is verified: 
    (1) $a=l(r_{Q_P})$ and $c=l(r_{T_p})$, (2) there exists
    $(x,y)\in P$ such that $x \in Q_{a-1}$ and $y \in T_{c-1}$.
  \item $(a,b;c,d)$ is called {\em $P$-right-maximal} if one of the
    following assertion is verified: (1) $b=r_Q-1$ and
    $d=r_T-1$, (2) there exists $(x,y)\in P$ such that $b+1 \in Q_x$
    and $d+1 \in T_{y}$.
  \end{enumerate}
\end{definition}

For example, in Fig.~\ref{fig:example}, let us consider the seed
$P=P^5$ ; then, $(4,12;2,8)$ is $P$-left-maximal as $(3,1)\in P$, 
$3$ is the root of $Q_{3=4-1}$ and its image $1$ in $T$ is the root of
$T_{1=2-1}$. Since $(13,9)$ is in $P$, $Q_{13}$ contains $12+1$ and $T_{9}$
contains $8+1$, assertion (2.2) of Definition~\ref{def:max} is also
verified and $(4,12;2,8)$ is $P$-right-maximal\footnote{Additional
  figures illustrating the definitions of this section are available
  in the Appendices.}.

\begin{definition}\em
\label{def:Fxy}
  Let $(x,y) \in B(P)$ for a seed $P$ between $Q$ and $T$. We define
  by $F(x,y)=\{(a_i,b_i;c_i,d_i)\}$ the set of all quadruples in
  $[l(x);x[^2 \times [l(y);y[^2$ that are both $P$-left-maximal and
          $P$-right-maximal and such that there is no border node of
          $P$ in $Q$ (resp. $T$) on the path from $b$ to $x$
          (resp. $d$ to $y$). We call this set the {\em chainable
            areas} of the border node $x$.
\end{definition}

For example, let us consider a pair $(x,y)$ in $L(P)$ such that $x$
and $y$ are not a leaf of respectively $Q$ and $T$, then $F(x,y)$
represents the couple of forests $\widehat{Q_x}$ and $\widehat{T_y}$,
$F(x,y)=\{(l(x),x-1;l(y),y-1)\}$.  In Fig.~\ref{fig:example}, with
$P=P^4$ and $(x,y)=(11,7)$, $F(x,y)=\{(8,10;5,6)\}$); if
$(x,y)=(14,11) \in B(P^5)-L(P^5)$, $F(x,y)=\{(0,1;0,0),(4,12;2,8)\}$. 


\begin{definition}\em
  \label{chainable_areas}
  The {\em chainable areas} of a seed $P$, denoted by $CA(P)$, is the
  union of the sets of quadruples $F(x,y)$ for all pairs $(x,y) \in
  B(P)$.
\end{definition}

\noindent{\em Notation.} For a seed $P$ (resp. chain $C$) and a
chainable area $(a,b;c,d)$, we say that $P \subset (a,b;c,d)$
(resp. $C \subset (a,b;c,d)$) if, for every $(x,y)\in P$
(resp. $(x,y)$ in a seed of $C$), $x \in Q[a,b]$ and $y \in T[c,d]$.

The following property is a relatively straightforward consequence of
the definitions of seeds and chainable areas.
\begin{property} \label{prpy:degree} 
  Given a seed $P$ between trees $Q$ and $T$, $|CA(P)|\leq 2\times
  |B(P)| + 1$.
\end{property}

The next property describes the structure of any chain between two
forests $Q[a,b]$ and $T[c,d]$ included in a set of $m$ seeds
$S=\{P^0,\ldots,P^{m-1}\}$. It is a direct consequence of the
constraints that define a valid mapping and the fact that seeds are
non-overlapping in a chain.

From now, for every $(x,y)$ of a seed $P^j$, we denote by $x^j$ the
unique node $y$ of $T$ associated with $x$ in $P^j$. We also denote by
$F_j(x)$ the set of quadruples $F(x,x^j)$ for the pair of nodes
$(x,x^j) \in P^j$.

\begin{property}
  \label{prpy:structure_chain}
    Let $P^j$ be the last seed of a chain $C$ included into two
    forests $Q[a,b]$ and $T[c,d]$.
    \begin{enumerate}
    \item
      $C$ can be decomposed into $|CA(P^j)|+2$ (possibly empty)
      distinct sub-chains: $P^j$ itself, $|CA(P^j)|$ chains: for each
      $(e,f;g,h)\in CA(P^j)$ a (possible empty) chain included into
      $Q[e,f]$ and $T[g,h]$ and a chain included into the forests 
      $Q[a,l(r_j)-1]$ and $T[c,l(r^j_j)-1]$.
    \item Moreover, $C$ is a chain of maximum score if all of
      its sub-chains described above are maximal.
   \end{enumerate}
\end{property}

Property~\ref{prpy:structure_chain}.2 naturally leads to a recursive
scheme to compute an optimal chain between two forests $Q[a,b]$ and
$T[c,d]$ that ends by the last seed of a set. If
$MCP'(Q[a,b],T[c,d],\{P^0\dots P^j\})$ is the score of a maximum chain
between $Q[a,b]$ and $T[c,d]$ and that contains $P^j$:
\begin{equation}
MCP'(Q[a,b],T[c,d],\{P^0\dots P^j\}) =  \label{eqn:seed_chain}
\end{equation}
\begin{equation*}
\left\{\begin{array}{lllr}
0 & \multicolumn{3}{r}{\text{if $P^j\not\subset (a,b;c,d)$,}} \\ \\
v(P^j)& \multicolumn{2}{l}{+ \displaystyle\sum_{(e,f;g,h)\in CA(P^j)} MCP(Q[e,f],T[g,h],\{P^0 \dots P^{j-1}\})} & \multirow{2}{*}{\text{\hspace*{2em}otherwise.}} \\
& + MCP(Q[a,l(r_j)-1],T[c,l(r^j_j)-1],\{P^0 \dots P^{j-1}\}) &\\
\end{array}\right.
\end{equation*}
and thus $MCP(Q,T,S)$ can be computed using $MCP'$ as
follow\footnote{We remind that the seeds are supposed to be sorted
  incrementally (see Remark~\ref{rem:seeds_order}).}: {\small
\begin{eqnarray}
MCP(Q[a,b],T[c,d],\{P^0\dots P^j\}) &=& \max_{i=0\dots j} MCP'(Q[a,b],T[c,d],\{P^0\dots P^i\}) \label{eqn:mcp_seeds}\\
MCP(Q,T,S)&=&MCP(Q[0,r_Q],T[0,r_T],S) \label{eqn:mcp}
\end{eqnarray}}

The main challenge in designing an algorithm for the MCP is then to
implement efficiently this recursive formula, that was already central
in the dynamic programming algorithm of~\cite{Heyne-lightweight}. In
Section~\ref{sec:2D}, we will rely on the fact that for every seed
$P^j$, $CA(P^j)$ and, for every border node $x$ of $P^j$, $F_j(x)$,
have been computed during a preprocessing phase. We discuss in
Appendix the issues related to this preprocessing and we show that it
can be done in $O(\|S\|)$ time and space.

\section{Algorithms for the Maximal Chaining Problem}
\label{sec:2D}


From now, we consider that we are given two ordered trees $Q$ and $T$,
a set $S=\{P^0,\ldots,P^{m-1}\}$ of seeds and a scoring score $v$ on
$S$. Furthermore, we assume that the score $v(j)$ of a seed $P^j$ can
be accessed in constant time and the seeds of $S$ are given as a list
$I$ of triples $(i,f,j)$ such that: (1) $i$ is the postfix number of
either the root of $P_Q^j$ or a border node of $P_Q^j$ (\emph{ie.}
$i\in B(P_Q^j)\cup r(P_Q^j)$) and (2) $f$ is a flag indicating if $i$
is either border ($f=0$) or root ($f=1$) for $P_Q^j$.  Thus if $i$ is
both in $B(P_Q^j)$ and the root of $P_Q^j$ then $i$ appears in two
distinct triples\footnote{Hence, we do not require as input the whole
  seeds mappings but just the borders and roots of the seeds, as it is
  usual when chaining seeds in sequences.}. Moreover, for a node $i$
in $Q$ belonging to a seed $P^j$, we assume that the corresponding
node in $T$, $i^j$ (or more precisely its postfix number in $T$) can
be accessed in constant time.  Finally, for every node $i$ in $Q$ and
$T$, its leftmost leaf $l(i)$ is also supposed to be accessed in
constant time.

As a preprocessing, $I$ is sorted in lexicographic order.
Thus, if a node is both in the border and root of $P^j$, it
first appears in $I$ as a border, then as a root.  This sorting can be
done in $O(||S||\log(||S||))$ time. In our algorithms, we visit
successively the elements of $I$ in increasing order, and a seed $P^j$
is said to be {\em processed} after its root has been processed
(\emph{i.e.} the current element of $I$ is greater than $(r_j,1,j)$ for the
order defined above).

In the following, we first introduce a {simple} but non optimal
algorithm to compute the MCP between $Q$ and $T$ which does not
require any special data structure.  In a second step, we will present
a more efficient method based on a simple modification of this
algorithm.

\subsection{A simple non optimal algorithm}
\label{sec:algo-imp1}


In order to compute in constant time the partial $MCP$ for any pair of
forests in $CA(P^j)$ as described in equation (\ref{eqn:seed_chain}),
we introduce a data structure $M$ indexed by quadruples of integers
$(a,b;c,d)$ defining the forests $Q[a,b]$ and $T[c,d]$. These
quadruples $(a,b;c,d)$ belong to a set $Y=Y_1 \cup Y_2 \cup Y_3$
defined as follows:
$$
Y_1=\bigcup_{j=0}^{m-1}CA(P^j),\
Y_2=\{(0,r_Q,0,r_T)\},$$
$$ 
Y_3=
\{(a,l(r_j)-1;c,l(r_j^j)-1)\text{ s.t. }  \exists (b,d) \text{
  s.t. } (a,b;c,d) \in Y_1 \cup Y_2
\text{ and } P^j\subset (a,b;c,d)\}
$$

In algorithm \ref{alg:second_version}, the function \texttt{Update} allows to
replace the value of $M[a,b,c,d]$ by a real number $w$ if $w$ is
greater than $M[a,b,c,d]$. We also use an array $V$ of $m$ integers to store the
intermediate quantities of $MCP^{'}$. 
The correctness of the algorithm relies on the following invariants
for the two data structures $V$ and $M$, that we prove later:
\begin{itemize}
\item[M1.] After $P^j$ has been processed, then $M[a,b,c,d]=MCP(Q[a,b],T[c,d],$\linebreak$\{P^0,\dots, P^j\})$ for every $(a,b;c,d)\in Y$.
\item[V1.] After $P^j$ has been processed, then $V[j]=MCP'(Q,T,\{P^0,\dots, P^j\})$.
\end{itemize}

\vfill\pagebreak

\begin{algorithm}\small
\caption{$MCP_1$: compute the score of a  maximal chain.}
\label{alg:second_version}
\begin{lstlisting}
for $j$ from $0$ to $m-1$ do $V[j]=v(j)$                         (*@\label{alg:second_version:init_v}@*)
foreach $(a,b;c,d) \in Y$ do $M[a,b,c,d]=0$                         (*@\label{alg:second_version:init_M}@*)
foreach $(i,f,j)$ in $I$ do 
  if $f=0$ then  # i.e. $(i,i^j) \in B(P^j)$                      (*@\label{alg:second_version:border_b}@*)
     foreach $(a,b;c,d) \in F_{j}(i)$ do $V[j]= V[j]+M[a,b,c,d]$ (*@\label{alg:second_version:border_e}\label{alg:second_version:border_v}@*)
  else # i.e. $f=1$ and $i$ is the root of $Q_{P^j}$, $i=r_j$              (*@\label{alg:second_version:root_b}@*)
    foreach $(a,b;c,d) \in Y_1\cup Y_2$ s.t. $P^j\subset(a,b;c,d)$ do (*@\label{alg:second_version:root_abcd}@*)
      Update $M[a,b,c,d]$ with $w=V[j]+M[a,l(r_j)-1,c,l({r^j_j})-1]$                               (*@\label{alg:second_version:update_m}@*)
         foreach $P^g \subset (r_j+1,b;r^j_j+1,d)$ do                   (*@\label{alg:second_version:root_rb}@*)
            Update $M[a,l(r_g)-1,c,l({r^g_g})-1]$ with $w$  
    $V[j]=V[j] + M[0,l(r_j)-1,0,l(r^j_j)-1]$                     (*@\label{alg:second_version:root_e}\label{alg:second_version:root_v}@*)
return $\max_{j} V[j]$
\end{lstlisting}
\end{algorithm}


\vspace*{-12mm}
\paragraph{Correctness of the algorithm.} 
Obviously, V1 implies that $\max_{j} V[j]$ contains the score of the
maximal chain (equations~(\ref{eqn:mcp_seeds}) and (\ref{eqn:mcp})).
Let us assume now that M1 is satisfied. If the seed $P^j$ has been processed,
then $V[j]$ contains the sum of $v(j)$
(line~\ref{alg:second_version:init_v}), the MCP scores of the
chainable areas of all its border nodes
(line~\ref{alg:second_version:border_v}) and the MCP score between
forests $Q(0,l(r_j)-1)$ and $T(0,l(r^j_j)-1)$
(line~\ref{alg:second_version:root_v}).  From
Property~\ref{prpy:structure_chain} and
(\ref{eqn:seed_chain}), $V[j]=MCP'(Q,T,\{P^0,\dots, P^j\})$ and
V1 is satisfied.

We prove M1 by induction. Initially, since no seed has been processed,
line~\ref{alg:second_version:init_M} ensures that M1 is satisfied. Now
let us assume that M1 is satisfied for all processed seeds
$\{P^0,\ldots,P^{j-1}\}$ and the input $(i,1,j)$ is being processed. If $P^j
\not\subset (a,b;c,d)$, then by induction, M1 is satisfied for
$M[a,b,c,d]$. Otherwise, the loop in
lines~\ref{alg:second_version:root_abcd} and
\ref{alg:second_version:update_m} ensures that M1 is satisfied for all
entries $M[a,b,c,d]$ such that $(a,b;c,d)\in Y_1 \cup Y_2$, as
$(a,l(r_j)-1;c,l({r^j_j})-1)$ does not contain $P^j$ ; thus by
induction M1 is satisfied for this index. Finally, the loop in
line~\ref{alg:second_version:root_rb} update all $(a,b;c,d)\in Y_3$ 
including $P^j$,and M1 is satisfied for all entries of
$M$.

\vspace*{-3mm}
\paragraph{Complexity analysis.} From Property~\ref{prpy:degree}, the space
required to encode the entries of $M$ indexed by $Y_1$ is in
$O(\|S\|)$. The space required to encode the entries of $M$ indexed by
$Y_3$ is in $O(m^2)$, as for every pair of seeds $P^i$ and $P^j$,
there is at most one chainable area of $CA(P^i)$ that contains $P^j$.

 We now address the worst-case time complexity. We do not factor the
 preprocessing required to compute the $F_j$ and $CA$ and we assume
 $I$ has been sorted in time $O(\|S\|\log(\|S\|))$. The amortized cost
 of
 lines~\ref{alg:second_version:border_b}--\ref{alg:second_version:border_e}
 is $O(\|S\|)$, as each chainable area is considered once, there are
 $O(\|S\|)$ such areas, and we assumed we can access them in amortized
 constant time. A naive implementation of
 lines~\ref{alg:second_version:root_b}--\ref{alg:second_version:root_v}
 would require $O(m^2\|S\|)$ operations: indeed, there are $m$
 iterations of the loop in line~\ref{alg:second_version:root_b}, the
 loop in line~\ref{alg:second_version:root_abcd} considers only
 entries indexed by $Y_1\cup Y_2$ (there are $O(\|S\|)$ such entries)
 and the loop on line~\ref{alg:second_version:root_rb} iterates $O(m)$
 times. However, we can notice that there are $O(m)$ entries
 $(a,b;c,d)\in Y_1\cup Y_2$ such that $P^j \subset (a,b;c,d)$, and it
 is possible to preprocess $I$ in time and space $O(m\|S\|)$ in such a
 way that the loop in line~\ref{alg:second_version:root_abcd} can be
 implemented to perform $O(m)$ iterations, leading to a total time
 complexity of $O(m\|S\|+m^3+\|S\|\log(\|S\|))$ (respectively for the
 preprocessing, the main algorithm and sorting the input).


\subsection{A more efficient algorithm}
\label{sec:algo-imp2}
The key ideas are to access less entries from $M$ (while maintaining
property M1 on the remaining entries though) and to complement $M$
with a data structure $R$ that can be queried in $O(\log(m))$ instead
of $O(1)$, but whose maintenance does not require a loop with $O(m^2)$
iterations. Formally, let $X=\{(a,c) \text{ s.t. } \exists (a,b;c,d)
\in Y_1\cup Y_2\}$ and $R$ be a data structure indexed by $X$ such
that for a given index $(a,c)\in X$, $R[a,c]$ is a set of pairs
$(j,s)$ where $j$ is the index of the seed $P^j$ and $s$ is the score
of the chain in $Q[a,r_j],T[c,r_j^j]$ that ends with $P^j$. Roughly, $M$ is used to access,
still in $O(1)$ time, the values $MCP(a,l(r_j)-1,c,l(r^j_j)-1,\{P^0
\dots P^{j-1}\})$ required to compute $MCP'$ in
equation~(\ref{eqn:seed_chain}) and $R[a,c]$ is used to access, in
time $O(\log(m))$, the scores of the best chains included in
$(Q[a,r_Q],T[c,r_T])$ (the values $MCP(Q[e,f],T[g,h],\{P^0 \dots
P^{j-1}\})$ in equation (\ref{eqn:seed_chain})) and replace the
entries $M[a,b,c,d]$ with $(a,b;c,d)\in Y_1\cup Y_2$ that were used in
the previous algorithm.


Finally, the algorithm iterates on a list of triples $J=I \bigcup
\left(\cup_{j=0}^{m-1}(l(r_j),-1,j)\right)$, sorted using the
lexicographic order than in the previous section, with the following
modification: if we have two seeds $P^j$ and $P^g$ with $g>j$ such
that $(l(r_j),l(r^j_j)) = (l(r_g),l(r^g_g))$ then only $(l(r_j),-1,j)$
occurs in $J$. This preprocessing requires $O(||S||\log(||S||))$ time.

\begin{algorithm}\small
\caption{$MCP_2(Q,T,S,v)$: compute a  maximal chaining from $S$.}
\label{alg:third_version}
\begin{lstlisting}[]
for $j$ from $0$ to $m-1$ do $V[j]=v(j)$                         (*@\label{alg:init_v}@*)
foreach $(a,b;c,d) \in Y_3$ do $M[a,b,c,d]=0$                        (*@\label{alg:init_M}@*)
foreach $(a,c) \in X$ do $R[a,c]=\emptyset$                          (*@\label{alg:init_R}@*)
foreach $(i,f,j)$ in $J$ do
  if $f=-1$ then # $i=l(r_j)$                                        (*@\label{alg:update_M0}@*)
    foreach $(a,c) \in X$ s.t. $a,c < l(r_j),l(r^j_j)$ do            (*@\label{alg:update_M1}@*)
      $M[a,l(r_j)-1,c,l({r^j_j})-1]$= value $s$ of the last $(y,s)$ of $R[a,c]$ s.t. $r_y^y < l({r^j_j})$(*@\label{alg:update_M2}@*)
  else if $f=0$ then #  $(i,i^j) \in B(P^j)$ (*@\label{alg:update_B0}@*)
    foreach $(a,b;c,d) \in F_{j}(i)$ do
      Add to $V[j]$ the value $s$ of the last entry $(y,s)$ of $R[a,c]$ s.t. $r_y^y \le d$ (*@\label{alg:update_V1}@*)
  else # $f=1$ and $i$ is the root of $Q_{P^j}$, $i=r_j$      (*@\label{alg:update_R1}@*)
    foreach $(a,c) \in X$ s.t. $a,c < l(r_j),l({r^j_j})$ do   (*@\label{alg:update_R2}@*)
      $w = V[j]+M[a,l(r_j)-1,c,l({r^j_j})-1]$                 (*@\label{alg:update_R3}@*)
      Insert entry $(j,w)$ into $R[a,c]$ and update $R[a,c]$ as follow: (*@\label{alg:update_R4}@*)
           Find the last entry $(y,s)$ s.t. $r_y^y < r_j^j$              (*@\label{alg:update_R5}@*)
           if $s < w$ then                                               (*@\label{alg:update_R6}@*)
                 Insert $(j,w)$ just after $(y,s)$ in $R[a,c]$ (*@\label{alg:update_R7}@*)
                 Remove from $R[a,c]$ all entries $(z,t)$ s.t. $r_j^j \leq r_z^z$ and $t < w$ (*@\label{alg:update_R8}@*)
    $V[j]=V[j] + M[0,l(r_j)-1,0,l(r^j_j)-1]$         (*@\label{alg:vj_left}@*)
return $\max_{j} V[j]$
\end{lstlisting}
\end{algorithm}

\noindent\paragraph{Correctness of the algorithm.} 
We consider the following invariants.
\vspace*{-1eM}\begin{itemize}
\item[M2.] After $P^j$ has been processed, then
  $M[a,b,c,d]=MCP(Q[a,b],T[c,d],$\linebreak$\{P^0,\dots,P^j\})$ for every
  $(a,b;c,d)\in Y_3$.
\item[V1.] After $P^j$ has been processed, then 
$V[j]=MCP'(Q,T, \{P^0,\dots,P^j\})$.
\item[R1.] After $P^j$ has been processed, then for all $(a,c) \in X$,
  $R[a,c]$ contains all $(y,s)$ that satisfies
\begin{itemize}
\item[a.] $y\le j \text{ and } s=MCP'(Q[a,r_y],T[c,r^y_y],\{P^0,\ldots,P^y\})$.
\item[b.] $\forall (z,t) \in R[a,c],~ r_z^z<r_y^y \Rightarrow t<s$.
\end{itemize}
\item[R2.] $\forall (a,c) \in X$, $R[a,c]$ is totally ordered as follows: $(y,s)<(z,t)$ iff $r_y^y < r_z^z$. 

\end{itemize}

We first assume that R1 and R2 are satisfied. As previously, if V1 is
satisfied, then the algorithm computes $MCP(Q,T,S)$.  The
initialization line~\ref{alg:init_v} ensures that $V[j]$ contains
$v(j)$. Next to prove V1 we only need to show that when we process a
border $i$ of a seed $P^j$, in line~\ref{alg:update_V1} we add to
$V[j]$ the best chains of each chainable area $(a,b;c,d)$ of the
border; it follows from (1) the fact that every seed $P^{j+e}$ with
$e>0$ does not belong to the forest $Q[a,b]$ (because $b<i\le
r_{j+e}$) and thus can not belong to a chain in the $(a,b;c,d)$ area,
(2) the fact that the score of this chain is present in $R[a,c]$ (from
R1) and (3) that fact it is the last entry $(y,s)$ such that $r_y^y
\le d$ (from R2).

M2 is similar to M1 but restricted to entries $M[a,b,c,d]$ such that
$(a,b;c,d)\in Y_3$. To check it is satisfied, we only need to focus on
line~\ref{alg:update_M2}, as it is the only line that updates $M$. For
entries $M[a,b,c,d]$ such that $a\geq l(r_j)$ or $c \geq l(r_j^j)$, then
$M[a,b,c,d]=0$ due to the initialisation in line~\ref{alg:init_v}. For
all other entries, M2 follows immediately from R1 and R2, using
argument similar to the previous ones.

Finally, we need to check that R1 and R2 are satisfied. First, as
previously, in the case where $a\ge l(r_j)$ or $c\ge l(r_j^j)$,
$R[a,c]=\emptyset$ which is ensured by the initialisation in
line~\ref{alg:init_R}. So we need only to consider the case where $a,c< l(r_j),l({r^j_j})$, that is handled in lines~\ref{alg:update_R1} to
\ref{alg:update_R8}. Every seed $P^y$ such that $y<j$ has already been
processed and $s=MCP'(Q[a,r_y],T[c,r^y_y],\{P^0,\ldots,P^y\})$ can not
be modified after $P^y$ has been processed, so
lines~\ref{alg:update_R2} and~\ref{alg:update_R3}, together with M2,
ensure that $(y,s)$ has been inserted into $R[a,c]$ previously, and
the same argument applies if $y=j$. Entries $(z,t)$ removed at
line~\ref{alg:update_R8} do not belong to any of these $(y,s)$, which
implies that R1.a and R1.b, and so R1, are satisfied. R2 is obviously
satisfied from the position where $(j,w)$ is inserted into $R[a,c]$ in
line~\ref{alg:update_R7}.

\paragraph{Complexity analysis.}
The space complexity is given by the space required for structures $M$
and $R$. $M$ requires a space in $O(m^2)$ as it is indexed by
$Y_3$. $R$ requires a space in $O(m\|S\|)$, as $|Y_1 \cup Y_2|\in
O(\|S\|)$ and for each seed $P^j$, an entry $(j,s)$ is inserted at
most once in each $R[a,c]$. All together, the space complexity is then
$O(m^2+m\|S\|)=O(m\|S\|)$.

We now describe the time complexity. First, note that following the
technique used for computing maximal chains in
sequence~\cite{Gusfield-algorithms,Joseph-determining,Ohlebusch-chaining},
the structures $R[l_Q,l_T]$ can be implemented using classical data
structures such as AVL or concatenable queues supporting query
requests, insertions, successor and, predecessor and deletions in a
set of $n$ totally ordered elements in $O(\log(n))$ worst-case time.

Now, we analyse the complexity of lines~\ref{alg:update_M0}
to~\ref{alg:update_M2}.  The loop of line~\ref{alg:update_M1} is
performed at most $O(m\|S\|)$ times and each iteration requires
$O(\log(m))$ in time (line~\ref{alg:update_M2}), which gives an
amortized time complexity of $O(m\|S\|\log(m))$.

Line~\ref{alg:update_V1} is applied at most once for each of the
$O(\|S\|)$ chainable area $F_j(i)$ (Property~\ref{prpy:degree}), and
each iteration requires $O(\log(m))$, which gives an $O(\|S\|\log(m))$
amortized time complexity.

Finally, we analyse the complexity of lines~\ref{alg:update_R1}
to~\ref{alg:vj_left}. First, we do not consider the operation in
line~\ref{alg:update_R8}. The number loop starting in
line~\ref{alg:update_R2} is performed in $O(m)$, and the complexity of
each loop is in $O(\|S\|)$. The cost of the operations performed
during each iteration is $O(\log(\|S\|))$ (lines~\ref{alg:update_R3}
and \ref{alg:update_R6} are both performed in $O(1)$ and
lines~\ref{alg:update_R4} and~\ref{alg:update_R5} in time
$O(\log(\|S\|)$). The total time complexity of this part, without
considering line~\ref{alg:update_R8}, is then $O(m\|S\|\log(\|S\|))$.
To complete the time complexity analysis, we show that the amortized
complexity of line~\ref{alg:update_R8} is in $O(m\|S\|)$. Indeed, it
follows from R2 that all entries removed in one step are consecutive
in the total order on $R[a,c]$ defined in R2. Hence, if one call to
line~\ref{alg:update_R8} removes $k$ elements from $R[a,c]$, it can be
done in $O((k+2)\log(m))$ time, as the successor of a given element
can be retrieved in $O(\log(m))$ time. As every element of $R$ is
removed at most once during the whole algorithm, this leads to an
amortized complexity of $O(m\|S\|\log(m))$ for
line~\ref{alg:update_R8}.
Alltogether, our algorithm solves computes $MCP(Q,T)$ in time
$O(m\|S\|\log(m))$, using standard data structures and after a
preprocessing in time $O(\|S\|\log(\|S\|))$ to compute the chainable
areas and to sort $J$.

\paragraph{Additional remarks.} 
  If we consider that $Q$ and $T$ are sequences, or, as described in
  Section~\ref{sec:preliminaries}, unary trees, then each of the two
  trees has a single leaf and each seed is unambiguously defined by
  its root and border, which implies that$\|S\|=m$. There is
  only one $R[a,c]$, as $a=c=0$, that contains $O(m)$ entries. Hence,
  all loops that were iterating on $R$ have now a single
  iteration, which reduces the time complexity by a factor $m$ to $O(\|S\|\log (m))=O(m\log(m))$.

  In the complexity analysis above, we followed the approach used for
  expressing the complexity of chaining in sequences, as we expressed
  the complexity only in terms of the size of the seeds. To express
  the complexity of our algorithm in terms of the size of $Q$ and
  $T$, a finer analysis of the data structure $R$ and of the number of
  different chainable areas leads to the following result: the
  worst-case space complexity of our algorithm is $O(|Q|^2|T|^2)$
  (similar to the algorithm of Heyne et al.), and its worst-case time
  complexity is in $O(\|S\|\log(\|S\|)+|Q||T|\log(|T|)(|Q||T|+m))$, to
  compare with the complexity of the Heyne et al algorithm that is in
  $O(\|S\|\log(\|S\|) + |Q|^2|T|^2(|Q||T|+m))$ (see details in the
  Appendix). This alternative complexity analysis is mostly of
  theoretical interest as in practice, for RNA analysis, one can
  expect that $m\ll|Q||T|$.  \renewcommand\baselinestretch{1}

\section{Conclusion}
\label{sec:conclusion}

The current paper describes algorithms to
solve chaining problems in ordered trees. With respect to
similar problems in sequences, these methods exhibit a linear factor
increase both in time and space.  Chains so obtained can be used to speed-up RNA
structure comparisons,
as illustrated in~\cite{Heyne-lightweight,Lozano-seeded}.

A natural question related to chaining problems, that, as far as we know, has
not been considered in the case of sequences, is to decide whether
a given seed $P$ of a set of seeds $S$ belongs to {\em
  any} optimal chains or not. However a trade-off between quality and speed
may have to be find. Indeed, identifying these \emph{always optimal} seeds
would ensure a good quality of the chains, whereas the high complexity
of these identifications would slow down the detection of similar
structures in a large database.


\smallskip
\noindent{\bf Acknowledgements.} Pacific Institute for Mathematical Sciences (PIMS, UMI CNRS 3069), Agence Nationale pour la Recherche project BRASERO (ANR-06-BLAN-0045), Natural Sciences and Engineering Research Council of Canada (NSERC), Multiscale Modeling of Plants associated team (INRIA).


~\vfill\pagebreak
\section*{Appendix: remarks on the dynamic programming algorithm of Heyne {\em et al.}}




This appendix proposes a comparison of the worst-case time complexity
between chaining algorithms proposed in the current paper and in
\cite{Heyne-lightweight}.

\subsection*{Time complexity of Heyne \emph{et al.} algorithm \cite{Heyne-lightweight}}
\label{sec:time-compl-heyne}

Heyne {\em et al.} algorithm \cite{Heyne-lightweight} considers pairs $(S_1,S_2)$ of
arc-annotated sequences, of respective length $n_1$ and $n_2$. 

This algorithm is based on processing {\em holes} in subsequences
corresponding to seeds, where a hole in an arc-annotated sequence is a
part of the subsequence spanned by a seed that does not belong to the
seed. In trees, holes correspond to the chainable areas of the border
nodes of a seed. The complexity of this algorithm is in $O(hn_1n_2)$
where $h$ is the number of different holes, $n_1$ and $n_2$
corresponds respectively to
the hole size in $S_1$ and $S_2$. Due to the
constraints on the definition of seeds proposed in \cite{Heyne-lightweight}
(connected nucleotides in RNA structure),
authors claim a time complexity of $O(n_1^2 n_2^2)$ as $h$ is bounded
by $O(n_1\times n_2)$ \cite{Heyne-lightweight}.

Actually, this complexity do not take into
account the time required to establish the holes and to sort them
(holes are treated in a specific order). Thus, the total time
complexity of this algorithm is $O(\|S\|\log \|S\| + h\times n_1\times
n_2)$ where $O(\|S\|)$ is the sum of seed sizes.

\paragraph{A worst-case complexity analysis}
 In the following, we
propose an analysis of the complexity of Heyne \emph{et al.} algorithm
\cite{Heyne-lightweight} in the case of more general seeds.

This algorithm uses dynamic programming tables
indexed by holes (in fact pairs of holes, one in each sequence defined
by a seed). Given a hole $h$ induced by a seed and defined by
the sequence $S_1[h^{L1},h^{R1}]$ in $S_1$ and the sequence $S_2[h^{L2},h^{R2}]$ in $S_2$,
$D^h(j,l)$ is the best chain included in $S_1[h^{L1},j]$ in $S_1$ and
$S_2[h^{L2},l]$ in $S_2$. Each hole is then processed independently from the
other ones (in an order ensuring required information have already
been computed), in order to fill the table $D^h$, using the following
dynamic programming equation:
$$ 
D^h[j,l] = \max\{D^h(j-1,l),D^h(j,l-1),\max_{\substack{\text{seed }P \in h \\\text{st.} P  \text{ ends on } j,l}}\{D^h(p-1,q-1)+S_P\}\}
$$ 
$$
S_P = v(P) + \sum_{h \in \text{Holes of $P$}} D^h(h^{R1},h^{R2})
$$
where, $p$ (resp. $q$) is the first base of seed $P$ in $S_1$
(resp. $S_2$) and $S_P$ is the score of the best chain included in the
subsequences spanned by $P$ in $S_1$ and $S_2$ and ended by $P$.

We can easily transpose this recurrence on trees, using article notation, as follow:
$$
D^{h=(a,b,c,d)}[j,l] = \max\left\{
\begin{array}[c]{l}
  D^h(j-1,l),\\
  D^h(j,l-1),\\
  \max_{\substack{\text{seed }P \subset (a,j,c,l) \\\text{st.}
      (r_P,r^P_P)=(j,l)}}\{D^h(l(r_P)-1,l(r^P_P)-1)+S_P\} 
\end{array}
\right \}
$$
$$
S_P = v(P) + \sum_{(a,b,c,d) \in CA(P)} D^{(a,b,c,d)}[b,d] = MCP'(Q_{r_P},T_{r^P_P},P^{0...P})
$$

First, let us remark that the computation of one dynamic programming matrix can 
be done in $O(n_1n_2+m)$ as the matrix has at most $n_1\times n_2$ entries and
the search of the seeds $P$ {\em which ends on $j,l$ } requires a pre-processing
in $O(m)$.

Thus, assuming that holes have already been computed, 
the total time complexity is $O(\|S\|\log(\|S\|) + h \times (n_1n_2+m) + \|S\|)$
(\emph{ie.} complexity of sorting of the holes plus the  computation
cost of $D^h$ plus the computation cost of $S_P$).

In \cite{Heyne-lightweight}, authors design seeds that are connected
nucleotides in the RNA secondary structure either by backbone bond
or base-pair bond. Hence, $h$ is bounded by $n_1n_2$ and the worst-case
time complexity is $O(\|S\|\log(\|S\|) + n_1n_2 \times (n_1n_2+m) + \|S\|)$.

If we impose seed nodes to be connected in the trees (and not in
RNA), which is a special case of our seeds but different from the
seeds developed by Heyne \emph{et al.}, the number of different holes would be
$O(n_1^2 n_2^2)$ in the worst case (all possible quadruplets
$(a,b,c,d)$). The overall complexity of the dynamic programming
algorithm then becomes in the worst case:
 $$O(\|S\|\log(\|S\|) + n_1^2n_2^2(n_1n_2+m)+\|S\|).$$

\subsection*{Time complexity of Algorithm 2}
\label{sec:time-compl-algo2}

To establish the worst-case complexity of  Algorithm~\ref{alg:third_version},
we have to study the cost of the algorithm for each $f$ values. To ease the reading,
we denote by $n_1$ the size of $Q$ and $n_2$ the size of $T$. Without loss of 
generality, we furthemore assume that $n_2\le n_1$.

Following invariants $R1$ and $R2$, each list of $R$ contains at most $min(m,n_2)$ elements, as there are
no $(y,s),(y',s') \in R[a,c]$ s.t. $r^y_y = r^{y'}_{y'}$, and $|X|\le min(\|S\|,n_1n_2)$.
Thus, in the worst-case, we have at most $O(n_1^2n_2^2)$ different chainable areas, $|R|=O(n_1n_2)$,
for all $(a,c)$: $|R[a,c]|=O(n_2)$ and $|X|=O(n_1n_2)$.
\begin{itemize}
\item[$f=-1$] line \ref{alg:update_M0}: Over the whole execution of the algorithm
each $M[a,l(r_j)-1,c,l(r^j_j)-1]$ is computed only once for all possible quadruplets as there
is no $(i,f,j)$, $(i',f',j')\in J$ such that $(l(r_j),l(r^j_j))=(l(r_{j'}),l(r^{j'}_{j'}))$.
Each computation require a search in $R[a,c]$ that can be done in $O(log(n_2))$. Thus, 
the total time complexity for this case is $O(n_1^2n_2^2 log(n_2))$.
\item[$f=0$] line \ref{alg:update_B0}: The computation line~\ref{alg:update_V1} can be store in a dedicated array $M'$ such
that the best chain of the area $(a,b,c,d)$ is computed only once. Thus, over all the execution of the algorithm, each 
different chainable area requires a search into a $R[a,c]$ and the total time complexity for this case is $O(\|S\|+n_1^2n_2^2\log(n_2))$.
\item[$f=1$] line \ref{alg:update_R1}: This case is run once peer seeds, so $O(m)$ times. Each run cost $O(n_1n_2\log(n_2))$ and the total 
time complexity is $O(mn_1n_2\log(n_2))$.
\end{itemize}

From above, we conclude that the worst-case time complexity of our algorithm is
\begin{eqnarray*}
&&O(\|S\|\log(\|S\|) + n_1^2n_2^2 log(n_2) + \|S\|+n_1^2n_2^2\log(n_2) +
mn_1n_2\log(n_2) )  \\
&&=
O(\|S\|\log(\|S\|) + n_1n_2\log(n_2)(n_1n_2+m)+\|S\|) \\
&&= O(\|S\|\log(\|S\|) + n_1n_2\log(n_2)(n_1n_2+m))
\end{eqnarray*}
which represents an improvement of the worst-case complexity of Heyne
et al. algorithm \cite{Heyne-lightweight}.

To conclude, we can merge the worst-case complexity analysis with the time complexity analysis of section~\ref{sec:algo-imp2}
leading to the following time complexity for Algorithm~\ref{alg:third_version}:

\begin{tabular}{llr}
$O($ & $\|S\| $ &computing the chainable areas \\
     & $+ \|S\| \log(\|S\|) $ & sorting the areas \\
     & $+ \min(m,n_1n_2) \times \min(\|S\|,n_1n_2) \times  log(\min(m,n_2)) $\hspace*{-2cm} & $f=-1$ case\\
     & $+ \|S\| + \min(\|S\|,n_1^2n_2^2) \times \log(\min(m,n_2))$ & $f=0$ case \\
     & $+m\times \min(\|S\|,n_1n_2)\log(\min(m,n_2))$ & $f=1$ case \\
\end{tabular}
as $|X|\le \min(\|S\|,n_1n_2)$ and $|R[a,c]| \le \min(m,n_2)$ for all $a,c$.

\newpage\section*{Appendix: additional figures}

\begin{figure}[!htb]
\begin{center}\includegraphics[width=0.8\linewidth]{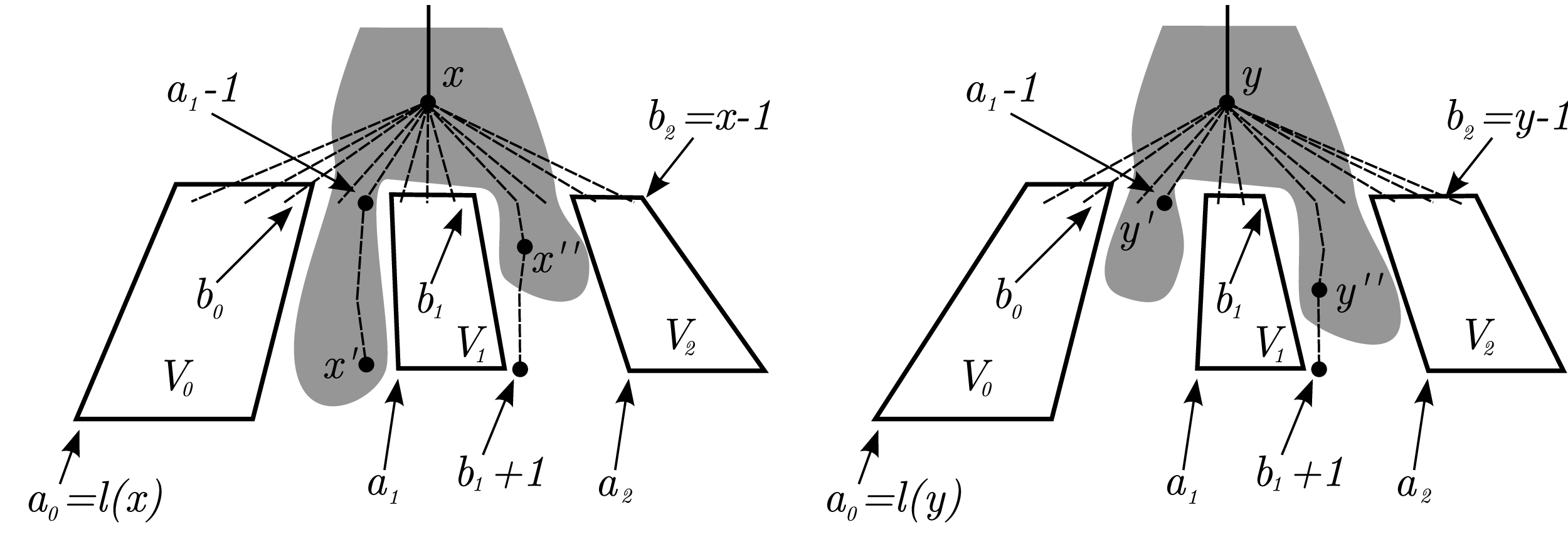}
 \caption{Illustration of Definition~\ref{def:Fxy} for a seed $P$ (the
   shaded zone) and $(x,y) \in B(P)-L(P)$:
   $F(x,y)=\{(a_0,b_0,c_0,c_1),(a_1,b_1,c_1,c_1),(a_2,b_2,c_2,c_2)\}$}
 \label{fig:below}
\end{center}
\end{figure}

\begin{figure}[!htb]
\begin{center}\includegraphics[width=0.9\linewidth]{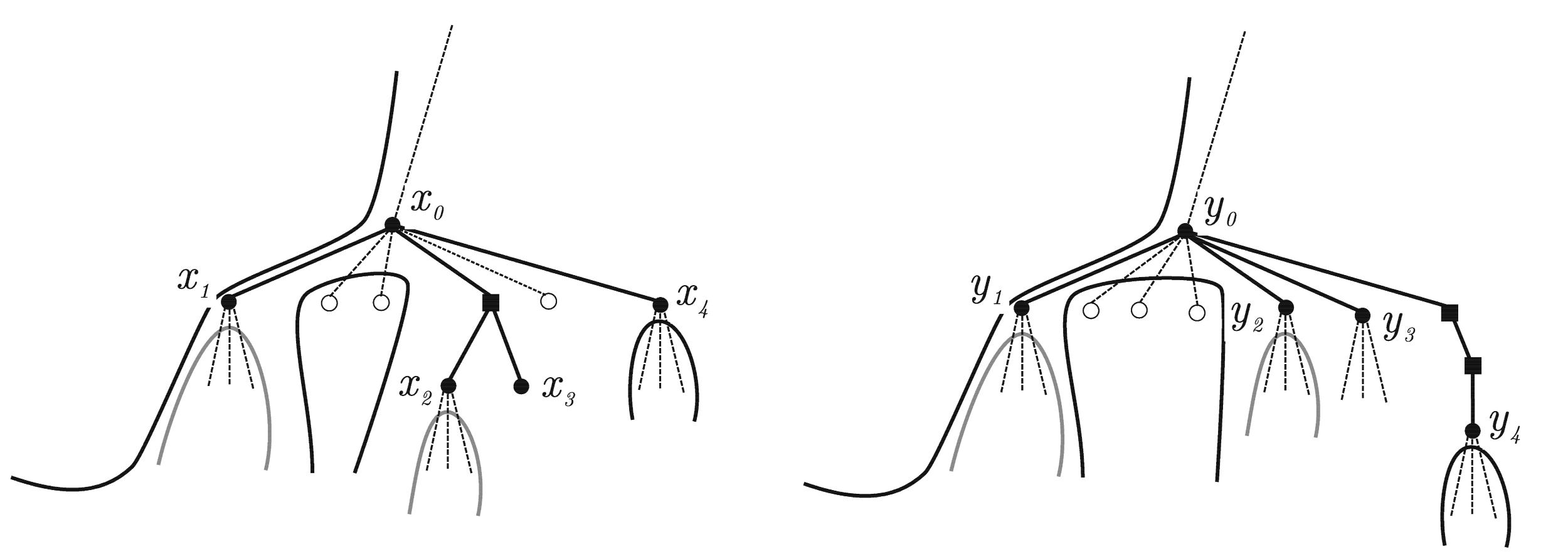}
\caption{Illustration of the notion of chainable areas of a seed of
  size 5: $P=\{(x_0,y_0),\ldots,(x_4,y_4)\}$ and $CA(P)$ contains $4$
  chainable areas each indicated by a different filling pattern. }
\label{fig:areas}
\end{center}
\end{figure}

\pagebreak\section*{Appendix: Computing $F(x,y)$ and families of seeds}

The cost of the computation of the chainable areas for the border
nodes of a seed depends of the nature of this seed. We describe here
an efficient algorithm that compute the $F(x,y)$.

Let $P$ be a seed between two trees $Q$ and $T$ and let $B(P)$ the set
of pairs of its border nodes. For each node $i$ of $Q$ and $T$ (in
fact, only required for the border nodes of $P$), we suppose that we
have access to the following informations in $O(1)$:
\begin{itemize}
\item $l(i)$: the leftmost leaf of  $i$.
\item $u(i)$: the node with the highest index such that $r(u(i))=r(i)$ where $r(i)$
  is the right most leaf of $i$.
\end{itemize}
The nodes $u(i)$ are often referred as rightmost roots of the tree.

In Algorithm~\ref{alg:chainable_area}, we use $B$ instead of
$B(P)$ and we assume that $B$ is an array of $k$ pairs of nodes on
$Q\times T$. For $0\le i<k$, $B[i]$ represents the $(i+1)^\text{th}$
pair of $B$ and $B[i]_Q$ is the node $Q$ of this pair and $B[i]_T$
is the node of $T$ of this pair.

Algorithm~\ref{alg:chainable_area} makes use of a stack of
pair of nodes called $Stack$.  $top(Stack)$ refers to the last element
inserted into $Stack$ and similarly to $B$, $top(Stack)_Q$ and
$top(Stack)_T$ are the node of $Q$ and node of $T$ of $top(Stack)$. We write
$push(Stack,(x,y))$ to add $(x,y)$ to the top of the stack and
$pop(Stack)$ remove the last element of the stack.

The algorithm that computes  $F(x,y)$ for all pair of border nodes
$(x,y)$ of $P$ is presented in Algorithm~\ref{alg:chainable_area}.

\paragraph{Description of the algorithm.} 
In the following,  a pair of border
node of $P$ is called shortly {\em a pair}.  Pairs are traversed incrementally according to their
postfix index. Hence, descendants are visited before parents. Remind
that a seed is a valid mapping so ancestral and order relations
between borders nodes are respected. Thus, if a border
node is a leaf in $P_Q$, it is also a leaf in $P_T$.

Before each insertion of a new chainable area into $F(x,y)$ we test
whether the area is non-empty or not (\emph{cf.}
lines~\ref{alg:ca:testleaf}, \ref{alg:ca:testright},
\ref{alg:ca:testmiddle}, \ref{alg:ca:testright} of
Algorithm~\ref{alg:chainable_area}).

Let us call the direct descendants of a pair $(x,y)$, the pairs $(x',y')
\in B(P)$ such that $x'$ is a descendant of $x$ (resp. $y'$ is a
descendant of $y$) and there is no border
node of $P$ in $Q$ (resp. $T$) between $x'$ and $x$ (resp.  $y'$ and
$y$).

Except for the last pair, each time a pair is visited, it is added
to the $Stack$ as it is necessarily the direct descendant of a none
visited pair.  Note that $Stack$ contains pairs sorted incrementally
by their postfix index.

Two cases must to be considered: (1) a pair is a pair of leafs in
$Q_P$, $T_P$ ($(x,y) \in L(P)$) and (2) a pair is not a pair of leafs
in $Q_P$,$T_P$ ($(x,y) \not\in
L(P)$). Lines~\ref{alg:ca:leaf}--\ref{alg:ca:leaf_end} correspond to
the first case and do not require additional explanations.

For the second case, the current pair $(x,y)$ necessarily has direct
descendants in $B(P)$. Those descendants have been visited (lower
postfix index) and thus are in $Stack$. The chainable area $(a,b,c,d)$
(possible empty) on the right of its rightmost direct descendant
$(x',y')$ ($a>x'$ and $c>y'$) and the chainable area (possible empty)
$(a,b,c,d)$ on the left of its leftmost direct descendant $(x'',y'')$
($b<x''$ and $d<y''$) require a particular treatment. The possible
chainable areas between two direct descendants are considered by the
loop on lines~\ref{alg:ca:loopmiddle}--\ref{alg:ca:loopmiddle_end}.
To compute these chainable areas, the following properties are used in
the algorithm: let $(x,y) \in B(P)$
\begin{enumerate}
\item Any chainable area $(a,b,c,d)$ of $(x,y)$ are such that $b$ and
  $d$ are children of $x$ and $y$ and $a$ and $c$ are the leftmost
  leafs of children of $x$ and $y$.
\item By definition, for chainable areas $(a,b,c,d)$ of $(x,y)$ except
  the one on the right of its rightmost descendant, $b$ and $d$ are
  such that $b+1$ and $d+1$ are the leftmost leafs of a direct
  descendant of $x$ and $y$.
\item For chainable areas $(a,b,c,d)$ of $(x,y)$ except the one on the
  left of its leftmost descendant, $a$ and $c$ are such that $a-1$ and
  $c-1$ are children of $x$ and $y$ and are either border nodes or
  ancestor of border nodes.
\end{enumerate}
As $Stack$ is sorted incrementally, the top of the stack contains the
rightmost direct descendant of current
pair. Lines~\ref{alg:ca:testright}--\ref{alg:ca:testright_end} compute
the chainable area on {\em the right} of this descendant. Then, loop
in lines~\ref{alg:ca:loopmiddle}--\ref{alg:ca:loopmiddle} compute the
area between the direct descendants using the above
properties. Finally,
lines~\ref{alg:ca:testleft}--~\ref{alg:ca:testleft_end} compute the
chainable area one {\em the left} the leftmost direct descendant (that
is the last pair $(x',y')$ in the $Stack$ such that $x'\ge l(x)$ and
$y'\ge l(y)$). Finally, remark that all direct descendant are now out
of the $Stack$ and are replaced by the current pair.

The time complexity of this algorithm is $O(|B(P)|)$ as we iterate of
all pair and each pair is added only once to the $Stack$.

Note that our algorithm is general as it applies to any sets of seeds
as defined in Definition~\ref{def:seed}. When considering restricted
families of seeds, it is possible to design simpler, while still
efficient, algorithms to compute the $F(x,y)$ and the chainable
areas. For example if one considers only {\em compact seeds},
i.e. seeds such that $B(P)=L(P)$ for every seed $P$, then for each
border $(x,y)$, $|F(x,y)|=1$ and the computation requires a time
linear in the number of seeds. A discussion on the issue of computing
chainable areas depending of the combinatorial nature of the
considered seeds will be added in a journal version of the current
work.

\begin{algorithm}\small
\caption{$F(x,y)$: compute the $F(x,y)$ for a seed $P$}
\label{alg:chainable_area}
\begin{lstlisting}
sort $B$ incrementally
foreach pair $(x,y)$ of $B$ do 
   $F(x,y)=\emptyset$
for $i$ from $0$ to $k-1$ do 
   if $i=0$ or $B[i-1]_Q<l(B[i]_Q)$ then #$B[i]$ is a pair of leafs in $Q_P$ and $T_P$ (*@\label{alg:ca:leaf}@*)
      if $l(B[i]_Q)\le B[i]_Q-1$ #$B[i]_Q$ is not a leaf in $Q$ (*@\label{alg:ca:testleaf}@*)
            and $l(B[i]_T)\le B[i]_T-1$ then #$B[i]_T$ is not a leaf in $T$
         $F(B[i])=(l(B[i]_Q),B[i]_Q-1,l(B[i]_T),B[i]_T-1)$ (*@\label{alg:ca:leaf_end}@*)
   else #$B[i]_Q$ has at least one descendant in $P$ that is the top of $Stack$ (*@\label{alg:ca:nonleaf}@*)
      # the rightest chainable area of $B[i]$:
      if $u(top(Stack)_Q)+1 \le B[i]_Q-1$ and $u(top(Stack)_T)+1 \le B[i]_T-1$ then (*@\label{alg:ca:testright}@*) 
         $F(B[i])=(u(top(Stack)_Q)+1,B[i]_Q-1,u(top(Stack)_T)+1,B[i]_T-1)$ (*@\label{alg:ca:testright_end}@*) 
      # the middle chainable areas of $B[i]$:
      $(x,y)=top(Stack)$; $pop(Stack)$
      while $Stack$ not empty and $top(Stack)_Q \ge l(B[i]_Q)$ do (*@\label{alg:ca:loopmiddle}@*)
         #insert the area between $(x,y)$ and $top(Stack)_Q$ if not empty
         if $u(top(Stack)_Q)+1 \le l(x)-1$ and $u(top(Stack)_T)+1 \le l(y)-1$ then (*@\label{alg:ca:testmiddle}@*)
            $F(B[i])+=(u(top(Stack)_Q)+1,l(x)-1,u(top(Stack)_T)+1,l(y)-1)$
         $(x,y)=top(Stack)$; $pop(Stack)$  (*@\label{alg:ca:loopmiddle_end}@*)
      # the leftest chainable area of $B[i]$
      if $l(x)-1 \ge l(B[i]_Q)$ and $l(y)-1 \ge l(B[i]_T)$ then (*@\label{alg:ca:testleft}@*)
         $F(B[i])+=(l(B[i]_Q),l(x)-1,l(B[i]_T),l(y)-1)$         (*@\label{alg:ca:testleft_end}@*)
   if $i<k-1$ then (*@\label{alg:ca:stack}@*)
      $push(Stack,())$
\end{lstlisting}
\end{algorithm}

\end{document}